\documentclass[conference]{IEEEtran}
\usepackage{graphicx,cite,amssymb,amsmath,psfrag,subfigure}
\usepackage{cite}
\usepackage{mathrsfs}
\usepackage[displaymath,mathlines]{lineno}
\usepackage{color}
\usepackage{tabulary}
\usepackage{multirow}
\usepackage{cases}
\usepackage{stfloats}
\usepackage{url}

\newtheorem{proposition}{Proposition}

\newtheorem{remark}{Remark}

\DeclareMathOperator{\tr}{\mathrm{tr}}
\DeclareMathOperator{\E}{\mathbb{E}}

\newcommand{\CG}[2]{\mathcal{CN}\left({#1},{#2}\right)}

\newcommand{\B}[1]{{\pmb{#1}}}

\newcommand{\Pu}{p_{\mathrm{u}}}
\newcommand{\Pd}{p_{\mathrm{d}}}

\newcommand{\tu}{\tau_{\mathrm{u}}}
\newcommand{\td}{\tau_{\mathrm{d}}}

\IEEEoverridecommandlockouts
\begin{document}
\title{
\hspace{4cm}\\[0cm]
     Massive MU-MIMO Downlink TDD Systems with Linear Precoding and Downlink Pilots}

\author{\IEEEauthorblockN{Hien Quoc Ngo\IEEEauthorrefmark{1}, Erik
G. Larsson\IEEEauthorrefmark{1}, and Thomas L.
Marzetta\IEEEauthorrefmark{2}}
\IEEEauthorblockA{\IEEEauthorrefmark{1}Department of Electrical Engineering (ISY)\\
Link\"{o}ping University, 581 83 Link\"{o}ping, Sweden}
\IEEEauthorblockA{\IEEEauthorrefmark{2}Bell Laboratories, Alcatel-Lucent\\
Murray Hill, NJ 07974, USA}
\thanks{
The work of H.~Q.\ Ngo and E.~G.\ Larsson was supported  in part
by the Swedish Research Council (VR), the Swedish Foundation for
Strategic Research (SSF), and ELLIIT. } }

\maketitle

\begin{abstract}
We consider a massive MU-MIMO downlink time-division duplex system
where a base station (BS) equipped with many antennas serves
several single-antenna users in the same time-frequency resource.
We assume that the BS uses linear precoding for the transmission.
To reliably decode the signals transmitted from the BS, each user
should have an estimate of its channel. In this work, we consider
an efficient channel estimation scheme to acquire CSI at each
user, called \emph{beamforming training} scheme. With the
beamforming training scheme, the BS precodes the pilot sequences
and forwards to all users. Then, based on the received pilots,
each user uses minimum mean-square error channel estimation to
estimate the effective channel gains. The channel estimation
overhead of this scheme does not depend on the number of BS
antennas, and is only proportional to the number of users. We then
derive a lower bound on the capacity for maximum-ratio
transmission and zero-forcing precoding techniques which enables
us to evaluate the spectral efficiency taking into account the
spectral efficiency loss associated with the transmission of the
downlink pilots. Comparing with previous work where each user uses
only the  statistical channel properties to decode the transmitted
signals, we see that the proposed beamforming training scheme is
preferable for moderate and low-mobility environments.
\end{abstract}

\IEEEpeerreviewmaketitle

\section{Introduction}

Recently, massive (or very large) multiuser multiple-input
multiple-output (MU-MIMO) systems have attracted a lot of
attention from both academia and industry
\cite{Mar:10:WCOM,LTEM:13:CM,NNSLZKL:13:CM,SYALMYZ:12:MobiCom}.
Massive MU-MIMO is a system where a base station (BS) equipped
with many antennas simultaneously serves several users in the same
frequency band. Owing to the large number of degrees-of-freedom
available for each user, massive MU-MIMO can provide a very high
data rate and communication reliability with simple linear
processing such as maximum-ratio combining (MRC) or zero-forcing
(ZF) on the uplink and maximum-ratio transmission (MRT) or ZF on
the downlink. At the same time, the radiated energy efficiency can
be significantly improved \cite{NLM:13:TCOM2}. Therefore, massive
MU-MIMO is considered as a promising technology for next
generations of cellular systems. In order to use the advantages
that massive MU-MIMO can offer, accurate channel state information
(CSI) is required at the BS and/or the users.

In small MU-MIMO systems where the number of BS antennas is
relatively small, typically, the BS can acquire an estimate of CSI
via feedback in frequency-division duplex (FDD) operation
\cite{KJC:11:COM}. More precisely, each user estimates the
channels based on the downlink training, and then it feeds back
its channel estimates to the BS through the reverse link. However,
in massive MU-MIMO systems, the number of BS antennas is very
large and channel estimation becomes challenging in FDD since the
number of downlink resources needed for pilots will be
proportional to the number of BS antennas. Also, the required
bandwidth for CSI feedback becomes very large. By contrast, in
time-division duplex (TDD) systems, owing to the channel
reciprocity, the BS can obtain CSI in open-loop directly from the
uplink training. The pilot transmission overhead is thus
proportional to the number of users which is typically much
smaller than the number of BS antennas. Therefore, CSI acquisition
at the BS via open-loop training under TDD operation is preferable
in massive MU-MIMO systems
\cite{Mar:10:WCOM,LTEM:13:CM,NNSLZKL:13:CM,HBD:11:ACCCC,JAMV:11:WCOM}.
With this CSI acquisition, in the uplink, the signals transmitted
from the users can be decoded by using these channel estimates. In
the downlink, the BS can use the channel estimates to precode the
transmit signals. However, the channel estimates are only
available at the BS. The user also should have an estimate of the
channel in order to reliably decode the transmitted signals in the
downlink. To acquire CSI at the users, a simple scheme is that the
BS sends the pilots to the users. Then, each user will estimate
the channel based on the received pilots. This is very inefficient
since the channel estimation overhead will be proportional to the
number of BS antennas. Therefore, the majority of the research on
these systems has assumed that the users do not have knowledge of
the CSI. More precisely, the signal is detected at each user by
only using the statistical properties of the channels
\cite{YM:13:JSAC,HBD:11:ACCCC,JAMV:11:WCOM}. Some work assumed
that the users have perfect CSI \cite{BHKD:13:DSP}. To the
authors' best knowledge, it has not been previously considered how
to efficiently acquire CSI at each user in the massive MU-MIMO
downlink.

In this paper, we propose a beamforming training scheme to acquire
estimates of the CSI  at each user. With this scheme, instead of
forwarding a long pilot sequence (whose length is proportional to
the number of BS antennas), the BS just beamforms a short pilot
sequence so that each user can estimate the effective channel gain
(the combination of the precoding vector and the channel gain).
The channel estimation overhead  of this scheme is only
proportional to the number of users. To evaluate the performance
of the proposed beamforming training scheme, we derive a lower
bound on the capacity of two specific linear precoding techniques,
namely MRT and ZF. Numerical results show that the beamforming
training scheme works very well in moderate and low-mobility
environments.

\textit{Notation:} We use upper (lower) bold letters to denote
matrices (vectors). The superscripts $T$, $\ast$, and $H$ stand
for the transpose, conjugate, and conjugate-transpose,
respectively. $\tr\left(\B{A}\right)$ denotes the trace of a
matrix $\B{A}$, and $\B{I}_{n}$ is the $n \times n$ identity
matrix. The expectation operator and the Euclidean norm are
denoted by $\mathbb{E}\left\{\cdot\right\}$ and $\| \cdot \|$,
respectively. Finally, we use $\B{z} \sim \CG{0}{\B{\Sigma}}$ to
denote a circularly symmetric complex Gaussian vector $\B{z}$ with
zero mean and covariance matrix $\B{\Sigma}$.

\section{System Model and Beamforming Training}

\begin{figure}[t]
\centerline{\includegraphics[width=0.48\textwidth]{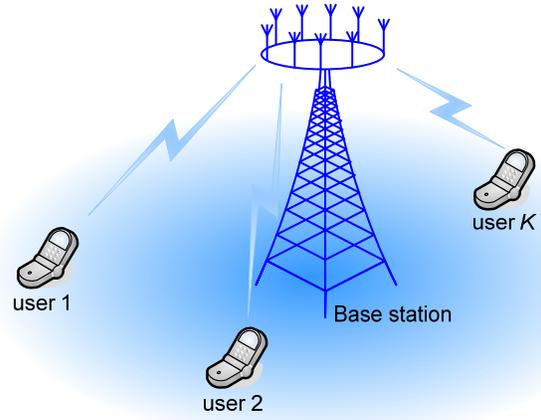}}
\caption{Massive MU-MIMO downlink system model.} \label{fig:1a}
\end{figure}

We consider the downlink transmission in a MU-MIMO system where a
BS equipped with $M$ antennas serves $K$ single-antenna users in
the same time-frequency resource, see Fig.~\ref{fig:1a}. Here, we
assume that $M \gg K$. We further assume that the BS uses linear
precoding techniques to process the signal before transmitting to
all users. This requires knowledge of CSI at the BS. We assume TDD
operation so that the channels on the uplink and downlink are
equal. The estimates of CSI are obtained from uplink training.

\subsection{Uplink Training}
Let $\tu$ be the number of symbols per coherence interval used
entirely for uplink pilots. All users simultaneously transmit
pilot sequences of length $\tu$ symbols. The pilot sequences of
$K$ users are pairwisely orthogonal. Therefore, it is required
that $\tu \geq K$.

 Denote by $\B{H} \in \mathbb{C}^{M\times
K}$ the channel matrix between the BS and the $K$ users. We assume
that elements of $\B{H}$ are i.i.d. Gaussian distributed with zero
mean and unit variance. Here, for the simplicity, we neglect the
effects of large-scale fading. Then, the minimum mean-square error
(MMSE) estimate of $\B{H}$ is given by \cite{KAY:93:Book}
\begin{align} \label{eq UL ParCSI 1}
    \hat{\B{H}}
    =
        \frac{\tu\Pu}{{\tu\Pu}+1}
        \B{H}
        +
        \frac{\sqrt{\tu\Pu}}{\tu\Pu +1} \B{N}_{\mathrm{u}}
\end{align}
where $\B{N}_{\mathrm{u}}$ is a Gaussian matrix with i.i.d.
$\CG{0}{1}$ entries, and $\Pu$ denotes the average transmit power
of each uplink pilot symbol. The channel matrix $\B{H}$ can be
decomposed as
\begin{align} \label{eq UL ParCSI 2}
    \B{H}
    =
        \hat{\B{H}}
        +
        \B{\mathcal{E}}
\end{align}
where $\B{\mathcal{E}}$ is the channel estimation error. Since we
use MMSE channel estimation, $\hat{\B{H}}$ and $\B{\mathcal{E}}$
are independent \cite{KAY:93:Book}. Furthermore, $\hat{\B{H}}$ has
i.i.d. $\CG{0}{\frac{\tu\Pu}{\tu\Pu+1}}$  elements, and
$\B{\mathcal{E}}$ has i.i.d. $\CG{0}{\frac{1}{\tu\Pu+1}}$
elements.

\subsection{Downlink Transmission}

Let $s_k$ be the symbol to be transmitted to the $k$th user, with
$\E\left\{|s_k|^2 \right\} = 1$. The BS uses the channel estimate
$\hat{\B{H}}$ to linearly precode the symbols, and then it
transmits the precoded signal vector to all users. Let $\B{W}\in
\mathbb{C}^{M\times K}$ be the linear precoding matrix which is a
function of the channel estimate $\hat{\B{H}}$. Then, the $M
\times 1$ transmit signal vector is given by
\begin{align} \label{eq DL IPCSI 1}
    \B{x}
    =
    \sqrt{\Pd}
    \B{W}
    \B{s}
\end{align}
where $\B{s} \triangleq \left[s_1 ~ s_2 ~ ... ~ s_K \right]^T$,
and $\Pd$ is the average transmit power at the BS. To satisfy the
power constraint at the BS, $\B{W}$ is chosen such as
$\E\left\{\left\|\B{x} \right\|^2 \right\}=  \Pd$, or equivalently
$\E\left\{\tr\left(\B{W}\B{W}^H \right) \right\}=  1$.

The vector of samples collectively received at the $K$ users is
given by
\begin{align} \label{Eq DL 4}
    \B{y}
    =
    \B{H}^T\B{x}
    +\B{n}
    =
    \sqrt{\Pd}
    \B{H}^T\B{W}\B{s}
    +\B{n}
\end{align}
where $\B{n}$ is a vector whose $k$th element, $n_k$, is the
additive noise at the $k$th user. We assume that $n_k \sim
\CG{0}{1}$. Define $a_{ki} \triangleq \B{h}_k^T \B{w}_i$, where
$\B{h}_i$ and $\B{w}_i$ are the $i$th columns of $\B{H}$ and
$\B{W}$, respectively. Then, the received signal at the $k$th user
can be written as
\begin{align}\label{Eq DL 5}
    y_k
    =
    \sqrt{\Pd}
    a_{kk} s_k
    +
    \sqrt{\Pd}\sum_{i\neq k}^K
    a_{ki} s_i
    +
    n_k.
\end{align}
%

\begin{remark}
Each user should have CSI to coherently detect the transmitted
signals. A simple way to acquire CSI is to use downlink pilots.
The channel estimate overhead will be proportional to $M$. In
massive MIMO, $M$ is large, so it is inefficient to estimate the
full channel matrix $\B{H}$ at each user using downlink pilots.
This is the reason for why most of previous studies assumed that
the users have only knowledge of the statistical properties of the
channels \cite{JAMV:11:WCOM,YM:13:JSAC}. More precisely, in
\cite{JAMV:11:WCOM,YM:13:JSAC}, the authors use
$\E\left\{a_{kk}\right\}$ to detect the transmitted signals. With
very large $M$, $a_{kk}$ becomes nearly deterministic. In this
case, using $\E\left\{a_{kk}\right\}$ for the signal detection is
good enough. However, for moderately large $M$, the users should
have CSI in order to reliably decode the transmitted signals. We
observe from \eqref{Eq DL 5} that to detect $s_k$, user $k$ does
not need the knowledge of $\B{H}$ (which has a dimension of $M
\times K$). Instead, user $k$ needs only to know $a_{kk}$ which is
a scalar value. Therefore, to acquire $a_{kk}$ at each user, we
can spend a small amount of the coherence interval on downlink
training. In the next section, we will provide more detail about
this proposed downlink \emph{ beamforming training} scheme to
estimate $a_{kk}$. With this scheme, the channel estimation
overhead is proportional to the number of users $K$.
\end{remark}

\subsection{Beamforming Training Scheme}

The BS beamforms the pilots. Then, the $k$th user will estimate
$a_{ki}$ by using the received pilots. Let $\B{S}_{\mathrm{p}} \in
\mathbb{C}^{K\times \tau_d}$ be the pilot matrix, where $\td$ is
the duration (in symbols) of the downlink training. The pilot
matrix is given by
\begin{align}\label{Eq BT 1}
    \B{S}_{\mathrm{p}}
    =
    \sqrt{\td \Pd}
    \B{\Phi}.
\end{align}
We assume that the rows of $\B{\Phi}$ are pairwisely orthonormal,
i.e., $\B{\Phi} \B{\Phi}^H = \B{I}_K$. This requires that $\tau_d
\geq K$.

The BS beamforms the pilot sequence using the precoding matrix
$\B{W}$. More precisely, the transmitted pilot matrix is
$\B{W}\B{S}_{\mathrm{p}}$. Then, the $K \times \td$ received pilot
matrix at the $K$ users is given by
\begin{align}\label{Eq BT 2}
    \B{Y}_{\mathrm{p}}^T
    &=
    \sqrt{\td \Pd}
    \B{H}^T\B{W}\B{\Phi}
    +\B{N}_{\mathrm{p}}^T.
\end{align}
where $\B{N}_{\mathrm{p}}$ is the AWGN matrix whose elements are i.i.d.
$\CG{0}{1}$. The received pilot matrix $\B{Y}_{\mathrm{p}}^T$ can
be represented by $\B{Y}_{\mathrm{p}}^T \B{\Phi}^H$ and
$\B{Y}_{\mathrm{p}}^T \B{\Phi}^H_{\perp}$, where
$\B{\Phi}^H_{\perp}$ is the orthogonal complement of $\B{\Phi}^H$, i.e., $\B{\Phi}^H_{\perp} = \B{I}_{\td} - \B{\Phi}^H \B{\Phi}$.
We can see that $\B{Y}_{\mathrm{p}}^T \B{\Phi}^H_{\perp}$ only
includes noise which is independent of $\B{Y}_{\mathrm{p}}^T
\B{\Phi}^H$. Thus, it is sufficient to use $\B{Y}_{\mathrm{p}}^T
\B{\Phi}^H$ for the channel estimation.  Let
\begin{align}\label{Eq BT 3}
    \tilde{\B{Y}}_{\mathrm{p}}^T
    \triangleq \B{Y}_{\mathrm{p}}^T \B{\Phi}^H
    &=
    \sqrt{\td \Pd}
    \B{H}^T\B{W}
    +\tilde{\B{N}}_{\mathrm{p}}^T
\end{align}
where $\tilde{\B{N}}_{\mathrm{p}}^T \triangleq
\B{N}_{\mathrm{p}}^T \B{\Phi}^H$ has i.i.d. $\CG{0}{1}$ elements.
From \eqref{Eq BT 3}, the $1 \times K$ received pilot vector at
user $k$ is given by
\begin{align}\label{Eq BT 4}
    \tilde{\B{y}}_{\mathrm{p},k}^T
    =
    \sqrt{\td \Pd}
    \B{h}_k^T\B{W}
    +\tilde{\B{n}}_{\mathrm{p},k}^T
    =
    \sqrt{\td \Pd}\B{a}_k^T +\tilde{\B{n}}_{\mathrm{p},k}^T
\end{align}
where $\B{a}_k \triangleq \left[a_{k1} ~ a_{k2} ~ ... ~ a_{kK}
\right]^T$, and $\tilde{\B{y}}_{\mathrm{p},k}$ and
$\tilde{\B{n}}_{\mathrm{p},k}$ are the $k$th columns of
$\tilde{\B{Y}}_{\mathrm{p}}$ and $\tilde{\B{N}}_\mathrm{p}$,
respectively.

From the received pilot $\tilde{\B{y}}_{\mathrm{p},k}^T$, user $k$
estimates $\B{a}_k$. Depending on the precoding matrix $\B{W}$,
the elements of $\B{a}_k$ can be correlated and hence, they should
be jointly estimated. However, here, for the simplicity of the
analysis, we estimate $a_{k1}, ..., a_{kK}$ independently, i.e.,
we use the $i$th element of $\tilde{\B{y}}_{p,k}$ to estimate
$a_{ki}$. In Section~\ref{Sec:Conclusion}, we show that estimating
the elements of $\B{a}_k$ jointly will not improve the system
performance much compared to the case where the elements of
$\B{a}_k$ are estimated independently. The MMSE channel estimate
of ${a}_{ki}$ is given by \cite{KAY:93:Book}
\begin{align}\label{Eq MMSE Est1}
    \hat{{a}}_{ki}
    =
    \E\left\{
        {a}_{ki}
    \right\}
    +
    \frac{\sqrt{\td \Pd}
    {\tt Var}\left({a}_{ki} \right)
        }{
        \td \Pd {\tt Var}\left({a}_{ki} \right)
        +
        1
        }
    \left(\tilde{{y}}_{\mathrm{p},ki} - \sqrt{\td \Pd}\E\left\{
        {a}_{ki}
    \right\}\right)
\end{align}
where ${\tt Var}\left({a}_{ki} \right)\triangleq
\E\left\{\left|{a}_{ki}- \E\left\{{a}_{ki}\right\} \right|^2
\right\}$, and $\tilde{{y}}_{\mathrm{p},ki}$ is the $i$th element
of $\tilde{\B{y}}_{\mathrm{p},k}$. Let ${\epsilon}_{ki}$ be the
channel estimation error. Then, the effective channel ${a}_{ki}$
can be decomposed as
\begin{align}\label{Eq MMSE Est2}
    a_{ki}
    =
    \hat{{a}}_{ki}
    +
    {\epsilon}_{ki}.
\end{align}
Note that, since we use MMSE  estimation, the estimate
$\hat{{a}}_{ki}$ and the estimation error ${\epsilon}_{ki}$ are
uncorrelated.

\section{Achievable Downlink Rate}

In this section, we derive a lower bound on the achievable
downlink rate for MRT and ZF precoding techniques, using the
proposed beamforming training scheme. To obtain these achievable
rates, we use the techniques of \cite{NLM:13:TCOM}.

User $k$ uses the channel estimate $\hat{\B{a}}_k$ in \eqref{Eq
MMSE Est1} to detect the transmitted signal $s_k$. Therefore, the
achievable downlink rate of the transmission from the BS to the
$k$th user is the mutual information between the unknown
transmitted signal $s_k$ and the observed received signal $y_k$
given by \eqref{Eq DL 5} and the known channel estimate
$\hat{\B{a}}_k = \left[\hat{a}_{k1} ~ ... ~ \hat{a}_{kK}
\right]^T$, i.e., $I\left(s_k;y_k,\hat{\B{a}}_k \right)$.

Following a similar methodology as in
\cite[Appendix~A]{NLM:13:TCOM}, we obtain a lower bound on the
achievable rate of the transmission from the BS to the $k$th user
as:
\begin{align}\label{Eq AR 1}
    R_k
    =
    \E\!\left\{\!
    \log_2\!\!
    \left(\!
        1 +
        \frac{
            \Pd \left| \hat{a}_{kk} \right|^2
            }{
             \Pd\sum\limits_{i=1}^K\E\left\{ \left|\epsilon_{ki}\right|^2\right\}
             +
             \Pd\sum\limits_{i\neq k}^K
              \left|\hat{a}_{ki}\right|^2
             +
             1
            }
    \!\right)
    \!\right\}.
\end{align}

We next simplify the capacity lower bound given by \eqref{Eq AR 1}
for two specific linear precoding techniques at the BS, namely,
 MRT and ZF.

\subsection{Maximum-Ratio Transmission}
With MRT, the precoding matrix $\B{W}$ is given by
\begin{align}\label{Eq MRT 1}
    \B{W}
    =
    \alpha_{\tt MRT}\hat{\B{H}}^{\ast}
\end{align}
where $\alpha_{\tt MRT}$ is a normalization constant chosen to
satisfy the transmit power constraint at the BS, i.e.,
$\E\left\{\tr\left(\B{W}\B{W}^H \right) \right\}=  1$. Hence,
\begin{align}\label{Eq MRT 2}
    \alpha_{\tt MRT}
    =
    \sqrt{
    \frac{1}{\E\left\{\tr\left(\hat{\B{H}}^{\ast}\hat{\B{H}}^T \right)
\right\}}
    }
    =
    \sqrt{
        \frac{
        \tu\Pu +1
        }{
        M K \tu\Pu
        }
        }.
\end{align}
\begin{proposition}\label{Pro1}
With MRT, the lower bound on the achievable rate given by
\eqref{Eq AR 1} becomes
\begin{align}\label{Eq MRT Pro1}
    R_k
    =
    \E\!\left\{\!
    \log_2\!
    \left(\!
        1 +
        \frac{
            \Pd \left| \hat{a}_{kk} \right|^2
            }{
             \frac{K \Pd }{\td\Pd +K}
             +
             \Pd\sum_{i\neq k}^K
              \left|\hat{a}_{ki}\right|^2
             +
             1
            }
    \!\right)
    \!\right\}
\end{align}
where
\begin{align}\label{Eq MRT Pro2}
    \hat{a}_{ki}
    =
    \frac{\sqrt{\td\Pd}}{\td\Pd+K}\tilde{\B{y}}_{p,ki}+ \frac{K}{\td\Pd+K}\sqrt{\frac{\tu\Pu M}{K \left(\tu\Pu +1
    \right)}}\delta_{ki}
\end{align}
where $\delta_{ki} = 1$ when $i=k$ and $0$ otherwise.
\begin{proof}
See Appendix~\ref{Appe 1}.
\end{proof}
\end{proposition}

\subsection{Zero-Forcing}

With ZF, the precoding matrix is
\begin{align}\label{Eq ZF 1}
    \B{W}
    =
    \alpha_{\tt ZF}\hat{\B{H}}^{\ast}
    \left(\hat{\B{H}}^{T}\hat{\B{H}}^{\ast}\right)^{-1}
\end{align}
where the normalization constant $\alpha_{\tt ZF}$ is  chosen to
satisfy the power constraint $\E\left\{\tr\left(\B{W}\B{W}^H
\right) \right\}=  1$, i.e., \cite{YM:13:JSAC}
\begin{align}\label{Eq ZF 2}
    \alpha_{\tt ZF}
    =
    \sqrt{
        \frac{
        \left(M-K\right)\tu\Pu
        }{
        K\left(\tu\Pu+1\right)
        }
        }.
\end{align}
\begin{proposition}\label{Pro2}
With ZF, the lower bound on the achievable rate given by \eqref{Eq
AR 1} becomes
\begin{align}\label{Eq ZF Pro1}
    R_k
    =
    \E\!\left\{\!
    \log_2\!
    \left(\!
        1 +
        \frac{
            \Pd \left| \hat{a}_{kk} \right|^2
            }{
             \frac{K \Pd }{\td\Pd +K\left(\tu\Pu+1 \right)}
             +
             \Pd\sum_{i\neq k}^K
              \left|\hat{a}_{ki}\right|^2
             +
             1
            }
    \!\right)
    \!\right\}
\end{align}
where
\begin{align}\label{Eq ZF Pro2}
    \hat{a}_{ki}
    &=
    \frac{\sqrt{\td\Pd}}{\td\Pd+K\left(\tu\Pu+1 \right)}\tilde{\B{y}}_{p,ki}
    \nonumber
    \\
    &+
    \frac{\sqrt{K\left(M-K\right)\tu\Pu\left(\tu\Pu+1 \right)}}{\td\Pd+K\left(\tu\Pu+1 \right)}
    \delta_{ki}.
\end{align}
\begin{proof}
See Appendix~\ref{Appe 2}.
\end{proof}
\end{proposition}

\begin{figure}[t]
\centerline{\includegraphics[width=0.45\textwidth]{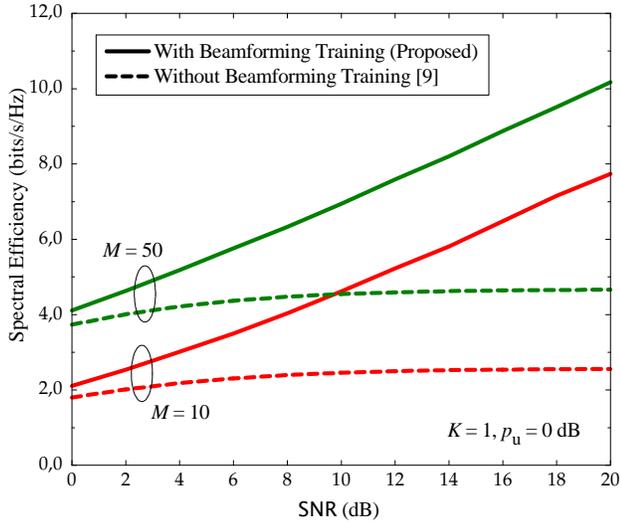}}
\caption{Spectral efficiency versus $\mathsf{SNR}$ for a
single-user setup ($K=1$, $\Pu=0$ dB, and $T = 200$).}
\label{fig:1}
\end{figure}

\begin{figure}[t]
\centerline{\includegraphics[width=0.45\textwidth]{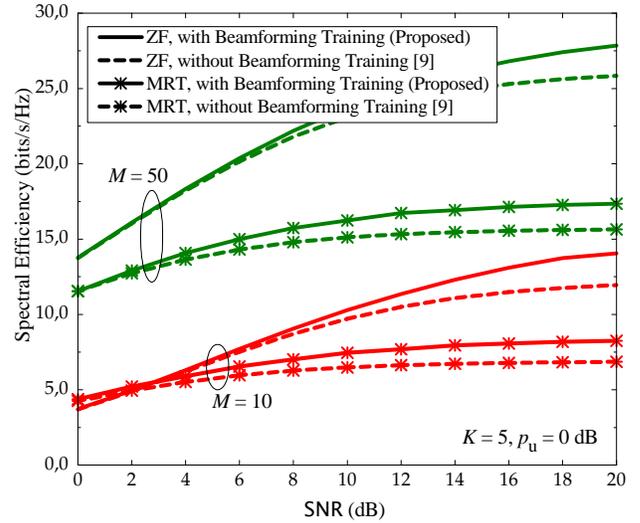}}
\caption{Spectral efficiency versus $\mathsf{SNR}$ for a multiuser
setup  ($K=5$, $\Pu=0$ dB, and $T = 200$).} \label{fig:2}
\end{figure}

\section{Numerical Results}
In this section, we illustrate the spectral efficiency performance
of the beamforming training scheme. The spectral efficiency is
defined as the sum-rate (in bits) per channel use. Let $T$ be the
length of the coherence interval (in symbols). During each
coherence interval, we spend $\tu$ symbols for uplink training and
$\td$ symbols for beamforming training. Therefore, the spectral
efficiency is given by
\begin{align}\label{Eq NR 1}
    \mathcal{S}_{\tt TB}
    =
    \frac{T-\tu-\td}{T}
    \sum_{k=1}^K R_k
\end{align}
where $R_k$ is given by \eqref{Eq MRT Pro1} for MRT, and \eqref{Eq
ZF Pro1} for ZF.

For comparison, we also consider the spectral efficiency for the
case that there is no beamforming training and that
$\E\left\{a_{kk} \right\}$ is used instead of $a_{kk}$ for the
detection \cite{YM:13:JSAC}. The spectral efficiency for this case
is given by \cite{YM:13:JSAC}
\begin{align}\label{Eq NR 2}
    \mathcal{S}_{0}
    \!=\!
\left\{\!\!\!
\begin{array}{l}
  \frac{T\!-\!\tu}{T}K \log_2\!\left(\!1+\frac{M}{K}\frac{\tu\Pu\Pd}{\left(\Pd+1 \right)\left(\tu\Pu+1 \right)} \!\right),  \text{for MRT} \\
  \frac{T\!-\!\tu}{T}K \log_2\!\left(\!1+\frac{M-K}{K}\frac{\tu\Pu\Pd}{\tu\Pu+\Pd+1 } \!\right),  \text{for ZF} \\
\end{array}%
\right.
\end{align}
In all examples, we choose $\tu=\td=K$ and $\Pu = 0$dB. We define
$\mathsf{SNR}\triangleq \Pd$.

We first consider a single-user setup ($K=1$). When $K=1$, the
performances MRT and ZF are the same. Fig.~\ref{fig:1} shows the
spectral efficiency versus $\mathsf{SNR}$ for different number of
BS antennas $M=10$ and $M=50$, at $T=200$ (e.g. $1$ms$\times
200$kHz). We can see that the beamforming training scheme
outperforms the case without beamforming training. The
performance gap increases significantly when the $\mathsf{SNR}$
increases. The reason is that, when $\mathsf{SNR}$ (or the
downlink power) increases, the channel estimate at each user is
more accurate and hence, the advantage of the beamforming training scheme grows.

Next, we consider a multiuser setup. Here, we choose the number of
users to be $K=5$. Fig.~\ref{fig:2} shows the spectral efficiency versus
$\mathsf{SNR}$ for the MRT and ZF precoders, at $M=10$, $M=50$, and
$T=200$. Again, the beamforming training offers an improved
performance. In addition, we can see that the beamforming training
with MRT precoding is more efficient than the beamforming training
with ZF precoding. This is due to the fact that, with ZF, the randomness of the effective channel gain $a_{kk}$ at the $k$th user is
smaller than the one with MRT (with ZF, $a_{kk}$ becomes deterministic when the BS has
perfect CSI) and hence, MRC has a higher advantage of using the channel estimate for the signal detection.

\begin{figure}[t]
\centerline{\includegraphics[width=0.45\textwidth]{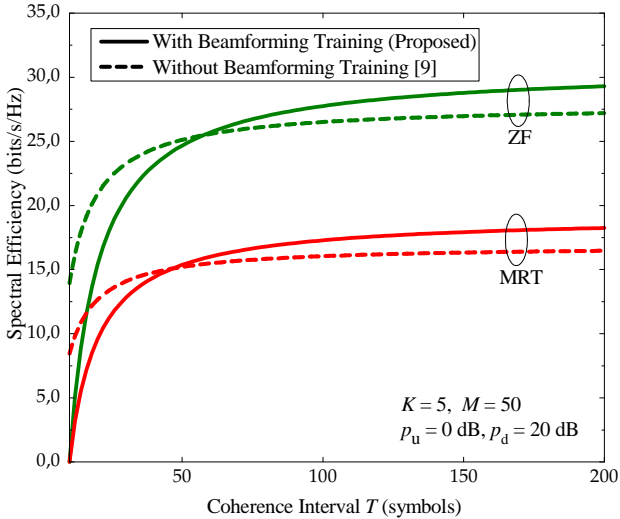}}
\caption{Spectral efficiency versus coherence interval for MRT and
ZF precoding ($M=50$, $K=5$, $\Pu=0$ dB, and $\Pd=20$ dB).}
\label{fig:3}
\end{figure}

Furthermore, we consider the effect of the length of the coherence interval on the
system performance of the beamforming training scheme.
Fig.~\ref{fig:3} shows the spectral efficiency versus the length
of the coherence interval $T$ at $M=50$, $K=5$, and $\Pd=20$ dB.
As expected, for short coherence intervals (in a high-mobility
environment), the training duration is relatively large
compared to the length of the coherence interval and hence, we should not use
 the beamforming training to estimate CSI at each user. At
moderate and large $T$, the training duration is relatively small
compared with the coherence interval. As a result, the beamforming
training scheme is preferable.

\begin{figure}[t]
\centerline{\includegraphics[width=0.45\textwidth]{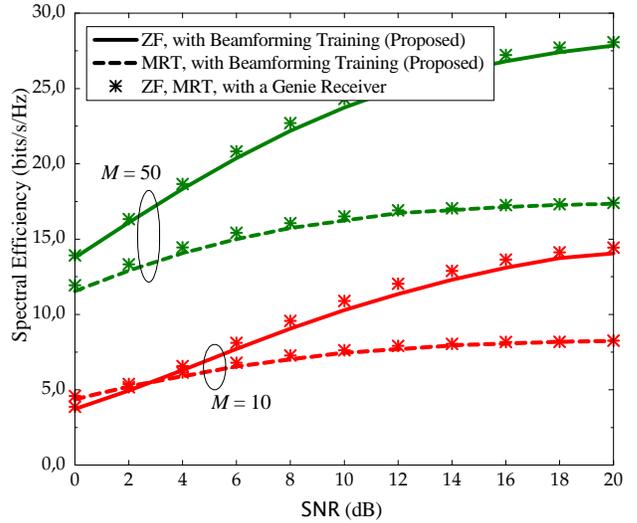}}
\caption{Spectral efficiency versus $\mathsf{SNR}$ with a genie
receiver ($K=5$, $\Pu=0$ dB, and $T=200$).} \label{fig:4}
\end{figure}

Finally, we consider  the spectral efficiency of our scheme but
with a genie receiver, i.e., we assume that the $k$th user can
estimate perfectly $\B{a}_k$ in the beamforming training phase.
For this case, the spectral efficiency is given by
\begin{align}\label{Eq NR 111}
    \mathcal{S}_{\tt G}
    \!=\!
    \frac{T\!-\!\tu\!-\!\td}{T}\!\!
    \sum_{k=1}^K\!
    \E\!\left\{\!
    \log_2\!\!
    \left(\!
        1 +
        \frac{
            \Pd \left| {a}_{kk} \right|^2
            }{
             \Pd\sum\limits_{i\neq k}^K
              \left|{a}_{ki}\right|^2
             +
             1
            }
    \!\right)
    \!\!\right\}.
\end{align}
Figure~\ref{fig:4} compares the spectral efficiency given by
\eqref{Eq AR 1}, where the $k$th user estimates the elements of
$\B{a}_k$ independently, with the one obtained by \eqref{Eq NR
111}, where we assume that there is a genie receiver at the $k$th
user. Here, we choose $K=5$ and $T=200$. We can see that
performance gap between two cases is very small. This implies that
estimating the elements of $\B{a}_k$ independently is fairly
reasonable.

\section{Conclusion and Future Work} \label{Sec:Conclusion}
In this paper, we proposed and analyzed a scheme to acquire CSI at
each user in the downlink of a MU-MIMO system, called beamforming
training scheme. With this scheme, the BS uses linear precoding
techniques to process the pilot sequence before sending it to the
users for the channel estimation. The channel estimation overhead
of this beamforming training scheme is small and does not depend
on the number of BS antennas. Therefore, it is suitable and
efficient for massive MU-MIMO systems. Furthermore, the down-link
pilots will add robustness to the beamforming process which
otherwise is dependent on the validity of the prior (Bayes)
assumptions.

\appendix
\subsection{Proof of Proposition~\ref{Pro1}}\label{Appe 1}
With MRT, we have that $ a_{ki}
    =
    \alpha_{\tt MRT}
    \B{h}_k^T
    \hat{\B{h}}_i^{\ast}$.
\begin{itemize}
\item Compute $\E\left\{a_{ki}\right\}$:

From \eqref{eq UL ParCSI 2}, we have
\begin{align}\label{Eq MRT Proof 1a}
    a_{ki}
    &=
    \alpha_{\tt MRT}
    \left(
        \hat{\B{h}}_k^T + {\B{\varepsilon}}_k^T
    \right)
    \hat{\B{h}}_i^{\ast}
    \nonumber
    \\
    &=
    \alpha_{\tt MRT} \hat{\B{h}}_k^T \hat{\B{h}}_i^{\ast}
    +
    \alpha_{\tt MRT}{\B{\varepsilon}}_k^T \hat{\B{h}}_i^{\ast}
\end{align}
where $\hat{\B{h}}_k$ and ${\B{\varepsilon}}_k$ are the $k$th
columns of $\hat{\B{H}}$ and $\B{\mathcal{E}}$, respectively.
Since $\hat{\B{\varepsilon}}_k$ and $\hat{\B{h}}_i^{\ast}$ are
uncorrelated with all $i, k = 1, ..., K$, we obtain
\begin{align}\label{Eq MRT Proof 2}
    \E\left\{a_{ki}\right\}
    &=
    \alpha_{\tt MRT}\E\left\{ \hat{\B{h}}_k^T \hat{\B{h}}_i^{\ast}\right\}
    \nonumber
    \\
    &=
    \left\{%
\begin{array}{l}
  0, ~ \text{if} ~ i\neq k \\
  \sqrt{\frac{\tu\Pu M}{K \left(\tu\Pu +1 \right)}}, ~ \text{if} ~ i=k \\
\end{array}%
\right.
\end{align}

\item Compute ${\tt Var}\left({a}_{ki} \right)$ for $i\neq k$:

From \eqref{Eq MRT Proof 1a} and \eqref{Eq MRT Proof 2}, we have
\begin{align}\label{Eq MRT Proof 3}
    {\tt Var}\left({a}_{ki} \right)
    &=
    \E\left\{\left|{a}_{ki}\right|^2 \right\}\nonumber
    \\
    &
    \mathop  = \limits^{(a)}
    \E\left\{\left|\alpha_{\tt MRT} \hat{\B{h}}_k^T \hat{\B{h}}_i^{\ast}\right|^2 \right\}
    +
    \E\left\{\left|\alpha_{\tt MRT}{\B{\varepsilon}}_k^T \hat{\B{h}}_i^{\ast}\right|^2 \right\}\nonumber
    \\
    &=
    \alpha_{\tt MRT}^2
    \left(
        \frac{
            \tu\Pu
            }{
            \tu\Pu +1
            }
    \right)^2 M
    +
    \alpha_{\tt MRT}^2
        \frac{
            \tu\Pu M
            }{
            \left(
            \tu\Pu +1
            \right)^2
            }\nonumber
    \\
    &=
    1/K
\end{align}
where $(a)$ is obtained by using the fact that $\hat{\B{h}}_k^T
\hat{\B{h}}_i^{\ast}$ and ${\B{\varepsilon}}_k^T
\hat{\B{h}}_i^{\ast}$ are uncorrelated.

\item Compute ${\tt Var}\left({a}_{kk} \right)$:

Similarly, we have
\begin{align}\label{Eq MRT Proof 4}
    {\tt Var}\left({a}_{kk} \right)
    &=
    \E\left\{\left|{a}_{kk}\right|^2 \right\}
    -\left|\E\left\{{a}_{kk} \right\}\right|^2.
\end{align}
From \eqref{Eq MRT Proof 1a}, we have
\begin{align}\label{Eq MRT Proof 5}
    \E\left\{\left|{a}_{kk}\right|^2 \right\}
    &=
    \alpha_{\tt MRT}^2\E\left\{\left\|\hat{\B{h}}_k\right\|^4 \right\}
    +
    \alpha_{\tt MRT}^2\E\left\{\left|{\B{\varepsilon}}_k^T \hat{\B{h}}_k^{\ast}\right|^2
    \right\}.
\end{align}
Using \cite[Lemma~2.9]{TV:04:FTCIT}, we obtain
\begin{align}\label{Eq MRT Proof 6}
    \E\left\{\left|{a}_{kk}\right|^2 \right\}
    &=
    \alpha_{\tt MRT}^2
    \left(
        \frac{
            \tu\Pu
            }{
            \tu\Pu +1
            }
    \right)^2 M\left(M+1\right)
    \nonumber
    \\
    &+
    \alpha_{\tt MRT}^2
        \frac{
            \tu\Pu
            }{
            \left(
            \tu\Pu +1
            \right)^2
            } M.
\end{align}
Substituting \eqref{Eq MRT Proof 2} and \eqref{Eq MRT Proof 6}
into \eqref{Eq MRT Proof 4}, we obtain
\begin{align}\label{Eq MRT Proof 7}
    {\tt Var}\left({a}_{kk} \right)
    &=
    1/K.
\end{align}
Substituting \eqref{Eq MRT Proof 2}, \eqref{Eq MRT Proof 3}, and
\eqref{Eq MRT Proof 7} into \eqref{Eq MMSE Est1}, we get \eqref{Eq
MRT Pro2}.

\item Compute $\E\left\{ \left|\epsilon_{ki}\right|^2\right\}$:

If $i\neq k$, from \eqref{Eq BT 4} and \eqref{Eq MRT Pro2}, we
have
\begin{align}\label{Eq MRT Proof 8}
    &\E\left\{ \left|\epsilon_{ki}\right|^2\right\}
    =
    \E\left\{ \left|a_{ki}-\hat{a}_{ki}\right|^2\right\}\nonumber
    \\
    &=
    \E\left\{ \left|\frac{K}{\td\Pd +K}a_{ki}-\frac{\sqrt{\td\Pd}}{\td\Pd +K}\tilde{n}_{\mathrm{p},ki}\right|^2\right\}
    \nonumber
    \\
    &=
    \left(\frac{K}{\td\Pd +K}\right)^2\E\left\{ \left|a_{ki}\right|^2\right\}
    +
    \frac{\td\Pd}{\left(\td\Pd +K\right)^2}
\end{align}
where $\tilde{n}_{\mathrm{p},ki}$ is the $i$th element of
$\tilde{\B{n}}_{\mathrm{p},k}$. Using \eqref{Eq MRT Proof 3}, we
obtain
\begin{align}\label{Eq MRT Proof 9}
    \E\left\{ \left|\epsilon_{ki}\right|^2\right\}
    =
    \frac{1}{\td\Pd +K}.
\end{align}
Similarly, we obtain  $\E\left\{
\left|\epsilon_{kk}\right|^2\right\}
    =
    \frac{1}{\td\Pd +K}$. Therefore, we arrive at the
    result in Proposition~\ref{Pro1}.

\end{itemize}

\subsection{Proof of Proposition~\ref{Pro2}}\label{Appe 2}
With ZF, we have that $ a_{ki}
    =
    \B{h}_k^T
    \B{w}_i$, where $\B{w}_i$ is the $i$th column of $\alpha_{\tt ZF}\hat{\B{H}}^{\ast}
    \left(\hat{\B{H}}^{T}\hat{\B{H}}^{\ast}\right)^{-1}$. Since $\hat{\B{H}}^T \B{W}= \alpha_{\tt
    ZF}\B{I}_K$, we have
\begin{align}\label{Eq ZF Proof 1}
    a_{ki}
    =
    \left(\hat{\B{h}}_k^T + {\B{\varepsilon}}_k^T\right) \B{w}_i
     =
     \alpha_{\tt ZF} \delta_{ki}
     + {\B{\varepsilon}}_k^T \B{w}_i.
\end{align}
Therefore,
\begin{align}\label{Eq ZF Proof 2}
    \E\left\{a_{ki}\right\}
    =
    \alpha_{\tt ZF} \delta_{ki}.
\end{align}
\begin{itemize}
\item Compute ${\tt Var}\left({a}_{ki} \right)$:

From \eqref{Eq ZF Proof 1} and \eqref{Eq ZF Proof 2}, we have
\begin{align}\label{Eq ZF Proof 3}
    {\tt Var}\left({a}_{ki} \right)
    &=
    \E\left\{\left|{\B{\varepsilon}}_k^T \B{w}_i\right|^2 \right\}
    =
    \frac{1}{\tu\Pu +1}
    \E\left\{\left\|\B{w}_i\right\|^2 \right\}
    \nonumber
    \\
    &=
    \frac{\alpha_{\tt ZF}^2}{\tu\Pu +1}
    \E\left\{\left[\left( \hat{\B{H}}^{T}\hat{\B{H}}^{\ast}\right)^{-1}\right]_{ii} \right\}
    \nonumber
    \\
    &=
    \frac{\alpha_{\tt ZF}^2}{\tu\Pu +1}\frac{1}{K}
    \E\left\{\tr\left[\left( \hat{\B{H}}^{T}\hat{\B{H}}^{\ast}\right)^{-1}\right]
    \right\}.
\end{align}
Using \cite[Lemma~2.10]{TV:04:FTCIT}, we obtain
\begin{align}\label{Eq ZF Proof 3b}
    {\tt Var}\left({a}_{ki} \right)
    =
    \frac{1}{K\left(\tu\Pu +1\right)}.
\end{align}
Substituting \eqref{Eq ZF Proof 2} and \eqref{Eq ZF Proof 3b} into
\eqref{Eq MMSE Est1}, we get \eqref{Eq ZF Pro2}.

\item Compute $\E\left\{ \left|\epsilon_{ki}\right|^2\right\}$:

If $i\neq k$, from \eqref{Eq BT 4} and \eqref{Eq ZF Pro2}, we have
\begin{align}\label{Eq ZF Proof 8}
    &\E\left\{ \left|\epsilon_{ki}\right|^2\right\}
    =
    \E\left\{ \left|a_{ki}-\hat{a}_{ki}\right|^2\right\}
    \nonumber
    \\
    &=
    \E\!\left\{\!
        \left|\!
            \frac{K\left(\tu\Pu+1 \right)a_{ki}
                }{
                \td\Pd +K\left(\tu\Pu+1 \right)
                }
                \!-\!
            \frac{
                    \sqrt{\td\Pd}\tilde{n}_{\mathrm{p},ki}
                        }{
                        \td\Pd +K\left(\tu\Pu+1 \right)
                        }
        \!\right|^2
    \!\right\}
    \nonumber
    \\
    &=
    \left(\frac{K\left(\tu\Pu+1 \right)}{\td\Pd +K\left(\tu\Pu+1 \right)}\right)^2\E\left\{ \left|a_{ki}\right|^2\right\}
    \nonumber
    \\
    &\hspace{3.9cm}+
    \frac{\td\Pd}{\left(\td\Pd +K\left(\tu\Pu+1 \right)\right)^2}
    \nonumber
    \\
    &=
    \frac{1}{\td\Pd +K\left(\tu\Pu+1 \right)}
\end{align}
where the last equality is obtained by using \eqref{Eq ZF Proof
3b}. Similarly, we obtain  $\E\left\{
\left|\epsilon_{kk}\right|^2\right\}
    =
    \frac{1}{\td\Pd +K\left(\tu\Pu+1 \right)}$. Therefore, we arrive at the
    result in Proposition~\ref{Pro2}.

\end{itemize}



\end{document}